\documentclass[pre,aps,twocolumn,showpacs]{revtex4}
\usepackage{graphics,graphicx,epsfig,color,subfigure}
\begin{document}
\title{Height distributions in competitive one-dimensional Kardar-Parisi-Zhang systems}
\author{Tiago J. Oliveira}
\email{tiago@ufv.br}
\affiliation{Departamento de F\'isica, Universidade Federal de Vi\c cosa, 36570-000, Vi\c cosa, MG, Brazil}
\date{\today}


\begin{abstract}
We study the competitive RSOS-BD model focusing on the validity of the Kardar-Parisi-Zhang (KPZ) ansatz $h(t) = v_{\infty} t + (\Gamma t)^{\beta} \chi$ and the universality of the height distributions (HDs) near the point where the model has Edwards-Wilkinson (EW) scaling. Using numerical simulations for long times, we show that the system is asymptotically KPZ, as expected, for values of the probability of the RSOS component very close to $p = p_c \approx 0.83$. Namely, the growth exponents converge to $\beta_{KPZ}=1/3$ and the HDs converge to the GOE Tracy-Widom distribution, however, the convergence seems to be faster in the last ones. While the EW-KPZ crossover appears in the roughness scaling in a broad range of probabilities $p$ around $p_c$, a Gaussian-GOE crossover is not observed in the HDs into the same interval, possibly being restricted to values of $p$ very close to $p_c$. These results improve recent ones reported by de Assis \textit{et al.} [Phys. Rev. E \textbf{86}, 051607 (2012)], where, based on smaller simulations, the KPZ scaling was argued to breakdown in broad range of $p$ around $p_c$. 
\end{abstract}

\maketitle

Surface growth is a subject of large interest in Statistical Physics, mainly due to pattern formation and emergency of universality in far from equilibrium systems, with applications ranging from biological growth to the production of modern thin film devices \cite{barabasi,krugl}. In stochastic modeling of the coarse-grained evolution of surfaces, the Kardar-Parisi-Zhang (KPZ) equation is one of the most important models, which reads \cite{KPZ}
 \begin{equation}
 \frac{\partial h (x,t)}{\partial t} = \nu \nabla^{2} h + \frac{\lambda}{2} (\nabla h)^{2} + \eta(x,t),
\label{eqKPZ}
\end{equation}
where $\nu$ accounts for the surface tension, $\lambda$ accounts for the growth velocity dependence on the local slope and $\eta$ is a white noise (with mean null and covariance $\left\langle \eta(x,t)\eta(x',t') \right\rangle = D \delta(x-x') \delta(t-t')$). Beyond growing interfaces, the KPZ class accomplishes several others non-equilibrium stochastic systems, such as directed polymers and exclusion processes \cite{barabasi,krug,krugrev}.

Almost one decade ago, the exact solution of some KPZ models in $d=1+1$ showed that the height evolves asymptotically in time as \cite{johansson,PraSpo}
\begin{equation}
h(t) = v_{\infty} t + (\Gamma t)^{\beta} \chi,
\label{eqAnsatz}
\end{equation}
where $v_{\infty}$ and $\Gamma$ are model dependent parameters, but the growth exponent $\beta=1/3$ and the asymptotic height distributions (HDs) of the random variable $\chi$ are universal. The HDs are associated with the Tracy-Widom distributions \cite{TW1} from the random matrix theory of Gaussian unitary ensemble (GUE) for curved growth and Gaussian orthogonal ensemble (GOE) for initial flat interfaces \cite{johansson,PraSpo}. Very recently, the universality of GUE and GOE distributions was proven experimentally \cite{TakeSano,TakeSano2}, numerically \cite{SidTiaSil1,SidTiaSil2,Takeuchi} and through the exact solution of the one-dimensional KPZ equation with some curved \cite{SasaSpo1,Amir} and flat \cite{CalaDoussal} geometries.

Such advances pose interesting new questions in the study of growing interfaces. For example, what happens to the HDs in competitive KPZ models ? Some works have analyzed the skewness of the HDs \cite{Talles1,Talles2} for short times, but the first work trying to answer this in the light of the Eq. \ref{eqAnsatz} was done very recently by de Assis \textit{et al.} \cite{Thiago}, which studied a competitive growth model (RSOS-BD) where particles are deposited with probabilities $p$ and $1-p$ following the restricted solid-on-solid (RSOS) and the ballistic deposition (BD) rules, respectively. Based on scaling exponents from the roughness dynamics and in the HDs properties, it was concluded in Ref. \cite{Thiago} that there exists three intervals of $p$ with different scaling behaviors: \textit{i}) $I_{B} \approx [0,0.75)$, where the system is in KPZ class, with BD-like surfaces; \textit{ii}) $I_{R} \approx (0.90,1]$, where again the KPZ class was found, but with surfaces closer to RSOS ones; and \textit{iii}) $I_{T}\approx (0.75,0.90)$, where the system is argued to not be in KPZ class. The growth exponent $\beta$ was said to be different from the KPZ one and the KPZ ansatz (Eq. \ref{eqAnsatz}) not to hold, i.e., it was not possible to measure the asymptotic growth velocity $v_{\infty}$ and the amplitude $\Gamma$. Furthermore, the skewness of the HDs was said to perform a continuous transition from a positive GOE value in $I_{B}$ to a negative GOE one in $I_{R}$, within the $I_{T}$ interval. Here, we perform simulations of the RSOS-BD model for times one order of magnitude larger than the ones studied in Ref. \cite{Thiago}. Our results show that, in $I_T$ interval, there is a long crossover from Edwards-Wilkinson \cite{EW} (EW - Eq. \ref{eqKPZ} with $\lambda=0$) to KPZ scaling. Thus, indicating that, asymptotically, a sudden KPZ-EW-KPZ transition must be expected at $p = p_c \approx 0.83$, where the system is in EW class. Another important result shown here is that the EW-KPZ crossover in the roughness scaling seems to be not accompanied by a crossover in the HDs from Gaussian (EW) distributions at short times to GOE (KPZ) ones asymptotically.

\begin{figure}[t]
\begin{center}
\includegraphics[width=8.5cm]{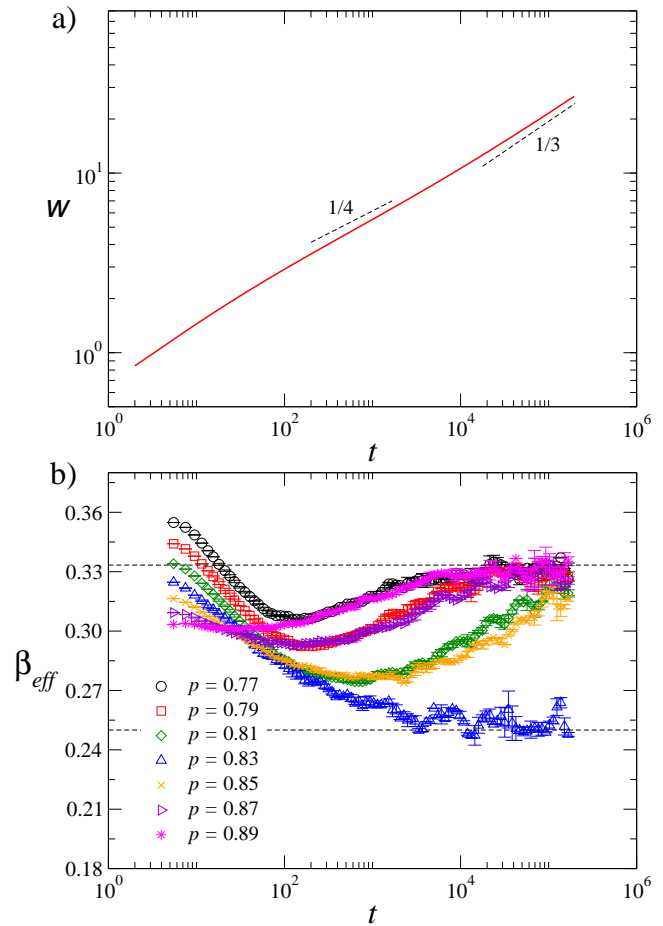}
\caption{(color online) (a) Roughness $w$ versus time $t$ for the probability $p=0.81$. (b) Effective growth exponents $\beta_{eff}$ as function of time for several values of $p$. The dashed lines indicates the slopes $\beta=1/4$ and $\beta=1/3$.} 
\label{fig1}
\end{center}
\end{figure}

It is well known, since a long time, that the competition of growth dynamics could produce crossovers in the roughness scaling. Generally, in models where particles are deposited accordingly to some KPZ rule, with probability $p$, and EW one with probability $1-p$, for small values of $p$ there is a crossover in the roughness evolution: In early times, the system seems to be in EW class, but in a characteristic time $t_{c}$ it crosses over to the KPZ scaling \cite{CrosEWKPZ} (see Fig. \ref{fig1}a). This occurs because the coefficient $\lambda$ in the KPZ equation (Eq. \ref{eqKPZ}) is a continuous function of $p$, which vanishes for $p=0$ \cite{Muraca,Flavio}. Therefore, for small $p$, $\lambda$ is also small and the laplacian-dominated growth produces the apparent initial EW regime. The same crossover appears when there is a competition between two models in the KPZ class with different signs of $\lambda$ in the corresponding KPZ equation. In this case, $\lambda=0$ for some value 	$p>0$ (see, for example, Ref. \cite{Tiago}). This is the case of the RSOS-BD model: Since $\lambda$ should be a continuous function of $p$ interpolating $\lambda_{BD}$ (positive) and $\lambda_{RSOS}$ (negative), there is a value of $p$ ($p_{c}\approx 0.83$) where $\lambda=0$ and the model has the scaling properties of the EW class. For any $p\neq p_c$, $\lambda \neq 0$ and the system must be KPZ asymptotically. \textit{There are not intermediate universality classes continuously interpolating the EW and KPZ ones}. However, as $\lambda \rightarrow 0$ ($p \rightarrow p_c$) the crossover time diverges with a very large exponent ($t_c \sim \lambda^{-4}$ \cite{Tiago}), thus, it is very difficult to observe the expected asymptotic KPZ behavior in numerical and/or actual experiments.

In order to analyze the existence, or possible breakdown, of the KPZ scaling in the RSOS-BD model close to $p_c$, we performed simulations of this model for $0.77\leq p\leq 0.89$ (the $I_T$ interval defined above). Out of this range of $p$'s, the KPZ scaling was clearly shown in Ref. \cite{Thiago}. Surfaces were grown to times up to $t=2\times 10^5$, on a substrate of size $L=2^{20}$ (one time unity corresponds to $L$ deposition attempts). This large system size is important to avoid saturation and finite-size effects in our analysis. For each $p$, averages over $100$ samples were done. It is important to notice that the RSOS particles are deposited at position $i$ if the constraint $(h_{i}-h_{i\pm 1})\leq 1$ is satisfied, as in the original definition of the RSOS-BD model by da Silva and Moreira \cite{Talles1}. Using the more restrictive RSOS condition $| h_{i}-h_{i\pm 1} |\leq 1$ the deposition of RSOS particles becomes very rare and the EW point $p_c$ becomes larger than $0.83$.

The surface roughness, or interface width $w$, is defined as the square root of the variance of the HDs. Fig. \ref{fig1}a shows the roughness evolution in time for $p=0.81$. After an initial transient, typical of BD models, an approximately EW scaling is found ($w\sim t^{1/4}$). However, for longer times, the system crosses over to the expected KPZ behavior ($w\sim t^{1/3}$). Obviously, calculating growth exponents from linear regressions of a log-log plot, as in Fig. \ref{fig1}a, into the (nonlinear) crossover region, we could wrongly conclude that they are different from the KPZ or EW ones. Indeed, a more reliable measure is provided by calculating effective growth exponents (local slopes in log-log plots of $w \times t$) as functions of time, as shown in Fig. \ref{fig1}b. For $p = 0.83  (\approx p_c)$ the exponent fluctuates around $\beta_{EW}=1/4$ for long times, expected for systems in EW class. For others probabilities $p$, the EW-KPZ crossover appears as a minimum in the effective exponents, which approaches $1/4$ when $p$ approaches $p_c$. After this initial scaling, the exponents converge to the KPZ one ($\beta_{KPZ}=1/3$) for all values of $p\neq p_c$ studied. As discussed before, the crossover time $t_c$, where the KPZ scaling begins to dominate, diverges when $p \rightarrow p_c$. In fact, in Fig. \ref{fig1}b we notice that the closer to $p_c$, the larger is the EW region and the larger are the times needed to attain the KPZ regime. For $p=0.81$ and $p=0.85$ the exponents did not reach the KPZ even for the large times studied, but they are very close to that value and still increasing, probably to that.

\begin{figure}[t]
\begin{center}
\includegraphics[width=8.15cm]{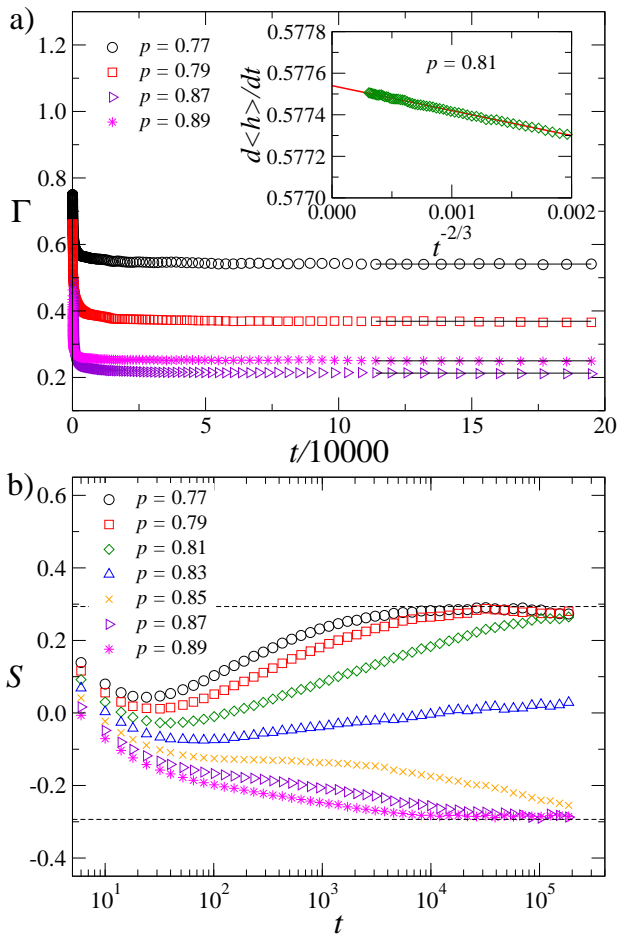}
\caption{(color online) (a) Non-universal parameter estimates: $\Gamma$ in main plot and $v_{\infty}$ in the inset. (b) Skewness of height distributions for several values of $p$ in the $I_T$ interval. The dashed lines indicates the expected KPZ GOE values $\pm 0.2935$.} 
\label{fig2}
\end{center}
\end{figure}

From Eq. \ref{eqAnsatz}, the non-universal parameter $\Gamma$ can be obtained from $\Gamma = \left[ \left\langle h^{2} \right\rangle_{c}/\left( \left\langle \chi^{2} \right\rangle_{c} t^{2/3}\right)\right]^{3/2}$, where $\left\langle Y^{2} \right\rangle_{c}$ is the variance of $Y$. $\Gamma$ becomes constant asymptotically, since $\left\langle h^{2} \right\rangle_{c} \sim t^{2/3}$ \cite{SidTiaSil2,TakeSano2,Thiago}. This scaling appears for long times in the interval $I_T$, thus it was concluded in Ref. \cite{Thiago} that it is not possible to obtain $\Gamma$ there. However, Fig. \ref{fig2}a shows clear plateaus of $\Gamma$ for $t\gtrsim 100000$ and $p < 0.81$ or $p >0.85$. Inside the interval $0.81 \leq p \leq 0.85$, the scaling $\left\langle h^{2} \right\rangle_{c} \sim t^{2/3}$ is not observed in the times studied here, as shown in Fig. \ref{fig1}b, possible due to the large crossover time. Nevertheless, these results suggest that it is always possible to determine $\Gamma$, except at $p=p_c$, where the KPZ ansatz (Eq. \ref{eqAnsatz}) does not hold (in other words, $\Gamma \equiv \lambda A^{2}/2$ is identically zero at $p_c$).

In the same way, the asymptotic growth velocity ($v_{\infty}$ in Eq. \ref{eqAnsatz}) could be obtained inside $I_T$, but for $p$ very close $p_c$ only at long times a linear behavior is found in a plot of $d\left\langle h\right\rangle /dt$ against $t^{-2/3}$. The inset of Fig. \ref{fig2}a shows this quantity for $p=0.81$ and $t \gtrsim 10000$. A very good linear behavior is found, allowing us to estimate $v_{\infty}$. We notice that, taking the complete time evolution of $d\left\langle h\right\rangle /dt$, i.e., including short times in the graph, a non-linear behavior is found, similar to the one shown in Fig. 3a of Ref. \cite{Thiago}. However, this initial non-linear behavior is due to crossover effects and should be discarded in the analysis.

Finally, in the height distributions (HDs) we still finding a strong evidence of KPZ behavior in the $I_T$ region, out of $p_c$. Fig. \ref{fig2}b shows the skewness evolution for several values of $p$ and, at long times, they present a good agreement with the Tracy-Widom GOE distribution ($S_{GOE} = 0.2935$), with exception of values of $p$ very close to $p_c$, where $S$ do not converged yet. For $p<p_c$ it converges to $+S_{GOE}$ because the system is BD dominated and this model has a positive $\lambda$, so that $S>0$. On the other hand, for $p>p_c$ the surface has RSOS-like features and, thus, $\lambda<0$ yielding $S<0$. For $p=0.83 (\approx p_c)$ the skewness is approximately null, as expected for EW systems, whose surfaces have up-down symmetry. We notice that plotting \textit{non-asymptotic} values of $S$ (for small times) as a function of $p$, as done in Ref. \cite{Thiago}, we could conclude that there exist a smooth transition between $+S_{GOE}$ and $-S_{GOE}$. However, Fig. \ref{fig2}b shows that increasing the times where $S$ is taken more abrupt transitions will appear, indicating that for $t\rightarrow\infty$ we must find a sudden discontinuous transition at $p=p_c$.

It is very important to notice in Fig. \ref{fig2}b that, besides the skewness of the HDs to have a minimum for $p<p_c$, it seems to not be related to an initial EW regime, in contrast to the minimum in effective exponents in Fig. \ref{fig1}b. For $p>p_c$, such minimal (in the absolute value of $S$) does not appear for the studied $p$'s. Therefore, the HDs seems do not have a clear crossover from EW to KPZ, i. e., they do not cross over from a Gaussian (EW) at small times to GOE (KPZ) at large ones. Thus, this suggest that the up-down symmetry is broken for any $p\neq p_c$ even at short times, where the roughness scaling presents an approximately EW behavior. However, it could be that for $p$ very close $p_c$ the crossover appears in HDs, but it is very difficult to observe because huge simulations times are needed to attain the KPZ regime. Anyway, this is different from the roughness scaling, where the crossover is present for a broad range of $p$.

In conclusion, we have performed large scale simulations of the competitive RSOS-BD model near the EW point $p=p_c$. For several values of $p$ close to $p_c$, our results are consistent with an asymptotic KPZ scaling, as expected. The growth exponents converge to $\beta=1/3$, but have a minimum associated with an initial EW regime. The non-universal parameters $v_{\infty}$ and $\Gamma$ from Eq. \ref{eqAnsatz} could be obtained near $p_c$, but as close to $p_c$ larger is the time needed to do this. Asymptotically, the HDs have a good agreement with the GOE Tracy-Widom distribution, but in short times it has absolute values of skewness $S$ smaller than $S_{GOE}$. All these results suggest that there is not a breakdown of the KPZ class inside the $I_T$ interval, in contrast to the proposal of Ref. \cite{Thiago}, but just a long EW-KPZ crossover that makes the KPZ properties to appear at very long times. Thus, for $t \rightarrow \infty$ we must expect a sudden KPZ-EW-KPZ transition at $p=p_c$. Another important result is that the EW-KPZ crossover in the roughness scaling seems do not be accompanied by a counterpart in the HDs, where an initial Gaussian (EW, $S=0$) regime was not observed for the $p's$ studied, suggesting that the up-down symmetry is severely broken for all $p\neq p_c$ at short (and long) times. This could lead the HDs to converge faster than the growth exponents to the asymptotic ones, as suggested by Figs. \ref{fig1}b and \ref{fig2}b. Therefore, the analysis of HDs could be more reliable than scaling exponents to determine universality in competitive systems.

We would like to thank F. D. A. Aar\~ao Reis for a critical reading of this manuscript and helpful discussions. We acknowledge the support from CNPq and FAPEMIG (brazilian agencies).

%~~~~~~~~~~~~~~~~~~~~~~~~~~~~~~~~~~~~~~~~~~~~~~~~~~~~~~~~~~~~~~~~~~~~~~~~~~~
%~~~~~~~~~~~~~~~~~~~  REFERENCES  ~~~~~~~~~~~~~~~~~~~~~~~~~~~~~~~~~~~~~~~~~~
%~~~~~~~~~~~~~~~~~~~~~~~~~~~~~~~~~~~~~~~~~~~~~~~~~~~~~~~~~~~~~~~~~~~~~~~~~~~

\end{document}